\documentclass[11pt,fleqn]{article}
\usepackage{amssymb}
\usepackage[latin1]{inputenc}
\usepackage{times}
\usepackage{mathptm}
\usepackage{epsfig}
\newcommand{\Section}[1]
{\section{#1}\setcounter{equation}{0}}

\newcommand{\abs}[1]{\left| #1 \right|}

\newcommand{\dagg}{{\scriptscriptstyle\dagger}}
\newcommand{\sign}{{\rm sign}}

\begin{document}
\mathindent 0mm
%%%%%%%%%%%%%%%%%%%%%%%%%%%%%%%%%%%%%%%%
\newpage\thispagestyle{empty} 
\begin{flushright} HD--TVP--98--1\end{flushright}
\vspace*{2.5cm}
\begin{center} 
{\Large Flow equations for band--matrices\\ \vspace*{0.4cm}}
\vskip1.5cm
Andreas Mielke\footnote[1]
{E--mail: mielke@hybrid.tphys.uni-heidelberg.de}\\ 
\vspace*{0.2cm}
Institut für Theoretische Physik,\\
Ruprecht--Karls--Universität,\\
Philosophenweg 19, \\
D-69120~Heidelberg, F.R.~Germany
\\
\vspace*{1cm}
{\large Dedicated to J. Zittartz on the occasion of his
$60^{\rm th}$ birthday.}

\vspace*{1.5cm}
\noindent
\today
\\[0.5cm]
Accepted for publication in European Physical Journal

\vspace*{1.5cm}
\noindent

{\bf Abstract}
\end{center}

\vspace*{0.2cm}\noindent
Continuous unitary transformations can be used to diagonalize
or approximately diagonalize a given Hamiltonian. 
In the last four years, this method has been applied to
a variety of models of condensed matter physics and
field theory. With a new generator for the continuous
unitary transformation proposed in this paper one can avoid
some of the problems of former applications. 
General properties of the new generator are derived. 
It turns out that the new generator is especially useful
for Hamiltonians with a banded structure. Two examples,
the Lipkin model, and the spin--boson model are discussed
in detail.

\vspace*{0.2cm}\noindent
PACS-Numbers: 03.65.-w, 02.70.Hm, 31.15.+q, 21.60.-n

\vspace*{2cm}
\newpage
\topskip 0cm
%%%%%%%%%%%%%%%%%%%%%%%%%%%%%%%%%%%%%%%%
\Section{Introduction}
%%%%%%%%%%%%%%%%%%%%%%%%%%%%%%%%%%%%%%%%
The diagonalization of a given Hamiltonian is one of
the important goals in any quantum mechanical problem.
Apart from few explicitly solved models it is
only possible approximatively or numerically.
Four years ago, Wegner \cite{Wegner94} 
proposed flow equations for Hamiltonians to
bring a given Hamiltonian closer to diagonalization. 
Technically, the approach is based on a continuous 
sequence of infinitesimal unitary transformations
applied to the Hamiltonian. The infinitesimal
unitary transformations are chosen so that the
off-diagonal matrix elements become smaller.
In principle the
flow equations proposed by Wegner finally yield a
diagonal matrix, except for the situation close to 
resonances, where few off-diagonal matrix elements
may still be large. This will be made more precise later.
Unfortunately it turned out that for a realistic Hamiltonian
in an infinite-dimensional Hilbert space, the situation may
be more difficult. Wegner discussed as an example 
interacting electrons in one dimension. The first
problem in such a system is, that during the flow
higher interactions are generated. Wegner solved this
problem by passing to an $n$-orbital model, where, in the
large $n$-limit, the flow equations can be closed. The
second problem is, that even then it is not possible
to diagonalize the Hamiltonian completely. Instead he
used a block--diagonalization. With this modified
approach he was able to solve the flow equations.

Later Wegner's flow equations have been applied to
various models of condensed matter physics. A
class of models where the flow equations yield
very accurate results are dissipative quantum systems
\cite{SK97a, SK97b}. It was also possible to treat
the electron--phonon problem with this method
\cite{LW96, Mielke97a}, for which one can obtain accurate
values for the transition temperature \cite{Mielke97b}. 
The reason why the approach is so successful is that
it provides a consistent renormalization scheme for
Hamiltonians. This has first been pointed out by G{\l}azek
and Wilson \cite{GW94}, who developed an essentially
equivalent method a few months later. Their method has been
used to treat problems from quantum chromo dynamics
\cite{Wilson94, Brisudova97}. 

Flow equations are a useful tool to treat systems
with various energy scales and to renormalize a given Hamiltonian,
but it is difficult to apply them to a finite matrix or to a 
Hamiltonian with a simple structure. Although it may be
possible to choose the transformation in such a way that some
of the higher interactions are not generated (this has been done
in the treatment of dissipative quantum systems in \cite{SK97a}),
one cannot avoid in general that the Hamiltonian looses its
initially simple structure. 
This point has been discussed in detail by Richter \cite{Richter97}.
He applied the flow equations to a simple model, the spin--boson model,
and used it to test various truncation schemes.
The aim of the present work is to propose
a different set of flow equations that has the property that 
a band diagonal matrix or Hamiltonian keeps its band diagonal structure
during the flow.

In the following section I introduce the new generator for
the continuous unitary transformation and I show that
it can be used to diagonalize a given Hamiltonian. The main
property of the new generator is that band diagonal Hamiltonians
remain band diagonal. I derive some properties of the new
flow equations. 

In section 3 and 4 I apply the flow equations to two simple
models: The Lipkin model and the spin--boson model. The
Lipkin model has a finite Hilbert space, the Hamiltonian
can be written in the form of two tridiagonal matrices. The
spin--boson model can as well be written as two tridiagonal
matrices, but the Hilbert space is infinite. The aim of this
paper is to show that the new set of flow equations is useful
if one wants to deal with simple finite or infinite matrices.
Therefore I only derive some results for the spectra of these
models, but I do not discuss the physics of these models in
detail. In both cases the reader may consult the references
for the physical background of the models. For the Lipkin model,
all results presented here are well known and have been obtained
by various other methods. For the spin--boson model I derive a
formula for higher eigenvalues that has been derived so far
only within a first order perturbational treatment. The new result
is that this formula has a much wider range of validity.

Section 5 contains the conclusions together with a critical
discussion of the possible range of applicability of the
new flow equations.

%%%%%%%%%%%%%%%%%%%%%%%%%%%%%%%%%%%%%%%%
\Section{Generalities}
%%%%%%%%%%%%%%%%%%%%%%%%%%%%%%%%%%%%%%%%
In this section I deal with a Hamiltonian that is given 
by a finite or infinite, real, symmetric matrix
\begin{equation}
  \label{Hamiltonian}
  H=(h_{nm}),\quad h_{nm}=h_{mn}.
\end{equation}
An extension to complex, hermitian or normal matrices is easily done.
In general, flow equations for a Hamiltonian are constructed by
a continuous unitary transformation written in a
differential form,
\begin{equation}
  \label{generalflow}
  \frac{dH}{d\ell}=[\eta,H]
\end{equation}
$\eta$ is the generator of the infinitesimal unitary
transformation, it is an anti-hermitian operator that
depends on $H$ and therefore implicitly on the flow parameter $\ell$. 
Wegner \cite{Wegner94}
proposed to choose $\eta=[H_d,H]$, where $H_d$ is the diagonal part
of the Hamiltonian. With this choice of $\eta$ one can show that
$\eta\rightarrow 0$ for $\ell\rightarrow\infty$. The flow equations
yield a final matrix with the property that $h_{nm}(h_{nn}-h_{mm})=0$.
This means that either the off-diagonal matrix elements $h_{nm}$ 
vanish or that for a finite $h_{nm}$ the difference of the 
corresponding diagonal matrix elements $h_{nn}-h_{mm}$ must vanish.
Thus the Hamiltonian is diagonalized except for some possible
resonances. As already mentioned in the introduction this choice
of $\eta$ has one disadvantage: If the initial matrix has
a simple structure, it looses this structure for finite $\ell$.
This becomes clear if one takes a band diagonal Hamiltonian, i.e.
\begin{equation}
  \label{banded}
  h_{nm}(\ell=0)=0\,\mbox{ if }\, \abs{n-m}>M
\end{equation}
Taking
\begin{equation}
  \eta=(\eta_{nm})
\end{equation}
one obtains flow equations for the matrix elements
\begin{equation}
  \frac{dh_{nm}}{d\ell}=\sum_k(\eta_{nk}h_{km}-h_{nk}\eta_{km})
\end{equation}
which in general do not conserve (\ref{banded}). 
Therefore I propose the new generator
\begin{equation} \label{eta}
  \eta_{nm}=-\eta_{mn}=\sign(n-m)h_{nm},\quad
  \eta_{nn}=0
\end{equation}
With this choice the
flow equations for the off-diagonal matrix elements are
\begin{equation}
  \label{flowofdiagonal}
  \frac{dh_{nm}}{d\ell}=-\sign(n-m)(h_{nn}-h_{mm})h_{nm}
  +\sum_{k\ne n,m}(\sign(n-k)+\sign(m-k))h_{nk}h_{km}.
\end{equation}
Due to the sum of the two $\sign$--functions the second term vanishes
if $\abs{n-m}>M$. This shows that the new generator
preserves the band diagonal structure. 
For the diagonal matrix elements one obtains
\begin{equation}
  \frac{dh_{nn}}{d\ell}=2\sum_{k\ne n}\sign(n-k)h_{nk}h_{kn}
\end{equation}
The main question is now whether the new choice of $\eta$ can be used to
diagonalize the Hamiltonian. This is indeed the case
for finite matrices ($1\le n\le N$) or semi--infinite matrices
($1\le n$). For the sum of the first $r$ diagonal matrix
elements one obtains the differential equation
\begin{equation}
  \frac{d}{d\ell}\sum_{n=1}^{r}h_{nn}=-2\sum_{n=1}^{r}\sum_{k>r}h_{nk}h_{kn}<0
\end{equation}
This quantity decays as a function of $\ell$.
I assume that $H$ is bounded from below. Then $\sum_{n=1}^{r}h_{nn}$
is bounded from below by the sum of lowest $r$ eigenvalues of $H$. 
Therefore its derivative must vanish in the limit $\ell\rightarrow\infty$,
i.e.
\begin{equation}
  \lim_{\ell\rightarrow\infty}h_{nk}h_{kn}=0
\end{equation}
Furthermore, if $h_{nm}$ tends to zero, one must have 
$\sign(n-m)(h_{nn}-h_{mm})<0$ for sufficiently large values
of $\ell$. Thus the diagonal matrix elements are ordered
for large $\ell$. 
 
Some properties of the flow equations with the new choice 
(\ref{eta}) of the generator are:
\begin{itemize}
\item The final matrix is diagonal, even if it contains degeneracies.
\item The matrix remains banded, $h_{nm}=0$ if $\abs{n-m}>M$.
\item $h_{nn}(\infty)\ge h_{mm}(\infty)$ for $n>m$ and for
irreducible matrices. If the matrix is reducible, each of the
irreducible blocks can be treated separately.
\item The asymptotic behaviour of $h_{nm}$ for large $\ell$
is known: $h_{nm}\propto \exp(-\abs{h_{nn}(\infty)-h_{mm}(\infty)}\ell)$.  
\end{itemize}
The last property follows directly from the flow equations. For large
values of $\ell$, the diagonal matrix elements are in the correct order
and no level crossings occur for larger values of $\ell$. The second
term in (\ref{flowofdiagonal}) falls of faster than 
$\propto \exp(-\abs{h_{nn}(\infty)-h_{mm}(\infty)}\ell)$ 
so that the asymptotic behaviour
is determined by the first term. 
For Wegner's choice of $\eta$, one cannot exclude non-vanishing 
off--diagonal matrix elements due to degeneracies of diagonal
matrix elements, and the asymptotic behaviour of the matrix
elements for large $\ell$ is less clear. 

Let me now discuss the application of the new flow equations to two examples.

\Section{Example 1: Lipkin model}

The Lipkin model \cite{Lipkin65}
is a toy model of nuclear physics that describes
in its simplest version two shells for the nucleons and an interaction
between nucleons in different shells. It serves as a standard example
for testing of various approximations. Recently Pirner and Friman
\cite{PF98} applied flow equations to this model. As usual
new interactions are generated and they used a suitable truncation
to close the flow equations.
They showed that for a large number of particles $N$ the truncated
flow equations yield the exact result whereas for small $N$ 
deviations occur. Furthermore the flow equations are only applicable
for small couplings. For larger couplings the model shows a transition
from a state with the same symmetry properties as the Hartree--Fock state
to a state with different properties \cite{DS90}.
I will come back to this point later.

A suitable representation for the Lipkin model is in terms of
pseudo-spin operators \cite{PF98}. 
\begin{equation}
  H(\ell=0)=\xi_0 J_z + V_0(J_+^2+J_-^2).
\end{equation}
$H$ commutes with $\hat{J}^2=J_z^2+\frac12(J_+J_-+J_-J_+)$ and 
$\hat{J}^2$ has the eigenvalues $J(J+1)$ as usual. The
pseudo-spin operators form the usual angular momentum algebra
\begin{equation}
  [J_z,J_\pm]=\pm J_\pm,\quad
  [J_+,J_-]=2J_z.
\end{equation}
It can easily be shown that in the basis where $\hat{J}^2$ and $J_z$
are diagonal the Hamiltonian decays into
two tridiagonal matrices.
\begin{equation}
  H=(h_{nm})_{n,m=0...J\mbox{ or }J-1\mbox{ or }J-1/2}
\end{equation}
The dimension of the matrices depends on $J$. If $2J$ is even, one
of the two matrices has dimension $J$, the other $J+1$, if $2J$
is odd, both matrices have the dimension $J+1/2$.
The matrix elements are
\begin{equation}
  h_{nn}=\epsilon_n,\quad
  h_{nn+1}=\delta_n,\quad
  h_{nm}=0\,\mbox{ if }\,\abs{n-m}>1
\end{equation}
with
\begin{equation}
  \epsilon_n(0)=\xi_0(-J+2n)
\end{equation}
\begin{equation}
  \delta_n(0)=V_0\sqrt{J(J+1)-(J-2n)(J-2n-1)}\sqrt{J(J+1)-(J-2n-1)(J-2n-2)}
\end{equation}
or
\begin{equation}
  \epsilon_n(0)=\xi_0(-J+2n+1)
\end{equation}
\begin{equation}
  \delta_n(0)=V_0\sqrt{J(J+1)-(J-2n-1)(J-2n-2)}\sqrt{J(J+1)-(J-2n-2)(J-2n-3)}.
\end{equation}
The flow equations are in both cases
\begin{equation} \label{flq_epsilon}
  \frac{d\epsilon_n}{d\ell}=-2\delta_n^2+2\delta_{n-1}^2
\end{equation}
\begin{equation} \label{flq_delta}
  \frac{d\delta_n}{d\ell}=-\delta_n(\epsilon_{n+1}-\epsilon_{n}).
\end{equation}
A first possibility is to solve these equations iteratively.
One can start with the {\it ansatz} $\epsilon^{(0)}_n(\ell)=\epsilon_n(0)$
and $\delta^{(0)}_n(\ell)=\delta_n(0)\exp(-2\xi_0\ell)$. Inserting these
expressions on the right hand side of the flow equations yields
a first iterative solution, which can again be used to obtain
the next iterative solution and so on. When one uses this
procedure it may be useful to write the flow equation for
$\delta_n$ in the form 
$\frac{d\ln\delta_n}{d\ell}=-\epsilon_{n+1}+\epsilon_{n}$.
This procedure reproduces simply perturbation theory, which
works well for small $V_0$ and not too large $J$ (i.e. not too
large particle numbers). 

A simple non-perturbative solution can be obtained in the limit of
large $J$; this corresponds to the limit of
a large particle number. The two different cases for the initial
conditions above yield
\begin{equation}
  \delta_n(0)^2-\delta_{n-1}(0)^2=32V_0^2J^2(n+\frac14)(1+O(1/J)),
\end{equation}
or
\begin{equation}
  \delta_n(0)^2-\delta_{n-1}(0)^2=32V_0^2J^2(n+\frac34)(1+O(1/J)).
\end{equation}
With the {\it ansatz}
\begin{equation}
  \epsilon_n(\ell)=a(\ell)n+b(\ell)
\end{equation}
the flow equations can be written as
\begin{equation}
  \delta_n(\ell)=f(\ell)\delta_n(0)
\end{equation}
where
\begin{equation}
  \frac{df}{d\ell}=-af,
\end{equation}
and
\begin{equation}
  \frac{da}{d\ell}=-64V_0^2J^2f^2.
\end{equation}
For the last two equations the quantity
\begin{equation}
  a^2-64V_0^2J^2f^2
\end{equation}
is conserved. Since $f\rightarrow 0$ for $\ell\rightarrow\infty$,
this yields directly
\begin{equation}
  a(\infty)=\sqrt{4\xi_0^2-64V_0^2J^2}
\end{equation}
This solution exists for $4JV_0<\xi_0$. 
Taking the first case from above one has
\begin{equation}
  \frac{db_1}{d\ell}=\frac14\frac{da}{d\ell}
\end{equation}
with the solution
\begin{equation}
  b_1(\infty)=-(J+\frac12)\xi_0+\frac14\sqrt{4\xi_0^2-64V_0^2J^2}
\end{equation}
For the second case one obtains
\begin{equation}
  \frac{db_2}{d\ell}=\frac34\frac{da}{d\ell}
\end{equation}
and
\begin{equation}
  b_2(\infty)=-(J+\frac12)\xi_0+\frac34\sqrt{4\xi_0^2-64V_0^2J^2}
\end{equation}
This yields directly the approximate spectrum
\begin{equation}
  \epsilon_{n1,2}=\sqrt{4\xi_0^2-64V_0^2J^2}(n+\frac12\pm\frac14)
  -(J+\frac12)\xi_0
\end{equation}
and the gap between the ground state and the first excited state
\begin{equation}
  \epsilon_{02}-\epsilon_{01}=\sqrt{\xi_0^2-16V_0^2J^2}.
\end{equation}
This result has been obtained by Pirner and Friman 
\cite{PF98} as well. It is also
well known from RPA. Nevertheless the above formulation of the
flow equations has certain advantages: 
\begin{enumerate}
\item It is quite easy to obtain perturbation theory using
the flow equations. With the conventional formulation of
the flow equations this in principle possible, but one has
to introduce many higher interactions if one wants to obtain
higher orders in perturbation theory. In the present formulation
the flow equations (\ref{flq_epsilon}, \ref{flq_delta}) are closed
and the iterative solution is easily constructed.
\item The formulation of the flow equations is not restricted
to $4JV_0<\xi_0$. Although the perturbative solution and
the approximate solution for large $J$ shown here are limited
to this regime, the flow equations (\ref{flq_epsilon}, \ref{flq_delta})
can be solved (at least numerically) for $4JV_0>\xi_0$ as well.
\item The flow equations in the present form may be used 
to derive a systematic $1/J$--expansion. To do this one has
to use a polynomial {\it ansatz} for $\epsilon_n$ as a function of $n$
instead of the linear {\it ansatz} above, and one has to take higher
orders in $\delta_n^2-\delta_{n-1}^2$ into account. 
\end{enumerate}
Since the aim of the present paper is only to show that the
new proposal for the continuous unitary transformation is
useful if one wants to treat band--diagonal Hamiltonians,
I do not follow the lines suggested in these points.

\Section{Example 2: Spin--boson model}

As a second example I discuss the spin--boson model described by 
the Hamiltonian
\begin{equation}
  \label{HSB}
  H(\ell=0)=-\frac{\Delta}2\sigma_x+\frac{\lambda}2\sigma_z(b+b^\dagg)
  +\omega b^\dagg b \, .
\end{equation}
It has a wide range of possible applications, especially in atomic physics
where the spin describes a two level atom that is coupled to a e.g.
laser field. Due to its long history there exists an enormous amount
of work that has already been published on this model, so that
it is impossible to review or cite all these papers. A good
overview may be found in the paper by Graham {\it et al}. \cite{GH84}. 
More recently
this model has been discussed in connection with quantum chaos
\cite{Cibils91, Cibils92}. Together with the usual flow equations
the model has been used to test
several approximation schemes \cite{Richter97}. It turned out that
the ground state and the low lying excited states
as well as dynamical properties can be calculated
very accurately using traditional flow equations. 
If one is interested in quantum chaos, an accurate knowledge
of high eigenvalues is necessary. 

The Hamiltonian (\ref{HSB}) can be written as two tridiagonal
infinite matrices. The flow equations are therefore the same
as for the Lipkin model,
\begin{equation}
  \frac{d\epsilon_n}{d\ell}=-2\delta_n^2+2\delta_{n-1}^2 ,
\end{equation}
and
\begin{equation}
  \frac{d\delta_n}{d\ell}=-\delta_n(\epsilon_{n+1}-\epsilon_{n}),
\end{equation}
but with different initial conditions:
\begin{equation}
  \delta_n(0)=\frac{\lambda}2\sqrt{n+1},
\end{equation}
\begin{equation}
  \epsilon_n(0)=n\omega \pm(-1)^n\frac{\Delta}2.
\end{equation}
It is very easy to solve these equations for $\Delta=0$. One obtains
\begin{equation}
  \epsilon_n=n\omega+\epsilon_{0},\quad
  \epsilon_{0}=-\frac{\lambda^2}{4\omega}(1-\exp(-2\omega\ell)),\quad
  \delta_{n}=\frac{\lambda}2\sqrt{n+1}\exp(-\omega\ell).
\end{equation}
In principle it is possible to use this solution to obtain an expansion
for small $\Delta$. A perturbative treatment for small $\Delta$ has been
given to first order by Graham {\it et al}. \cite{GH84}, and the flow equations
yield the same result. Therefore I will not reproduce this solution here.
A second possibility is an iterative solution for small $\lambda$. It
yields a perturbative solution valid for small $n$. Similarly 
ordinary perturbation theory is valid only if $n$ is small
($n\ll (\omega\pm\Delta)^2/\lambda^2$).
Instead I try to obtain an asymptotic expression for $\epsilon_n$ 
that is vaild for large $n$.
To do this, I make the following {\it ansatz}
\begin{equation}
  \epsilon_n=n\omega-\frac{\lambda^2}{4\omega}(1-\exp(-2\omega\ell))
  \pm (-1)^n\frac{\Delta}2f_n(\ell),
\end{equation}
and
\begin{equation}
  \delta_n^2=\frac{\lambda^2}4(n+1)\exp(-2\omega\ell)
  \pm (-1)^n\frac{\Delta}2g_n(\ell).
\end{equation}
This {\it ansatz} yields flow equations for $f_n$ and $g_n$
\begin{equation}
  \frac{df_n}{d\ell}=-2(g_n+g_{n-1})
\end{equation}
\begin{equation}
  \frac{dg_n}{d\ell}=\frac{\lambda^2}2(n+1)\exp(-2\omega\ell)(f_{n+1}+f_n)
  -2\omega g_n
  \mp \Delta (-1)^ng_n(f_{n+1}+f_n)
\end{equation}
It is now useful to introduce the new variable
\begin{equation}
  x=1-\exp(-2\omega\ell)
\end{equation}
instead of the flow parameter $\ell$. In the following I take
$f_n$ and $g_n$ as functions of $x$. The flow equations are
rewritten as
\begin{equation}
  \omega(1-x)\frac{df_n}{dx}=-g_n-g_{n-1},
\end{equation}
\begin{equation}
  2\omega(1-x)\frac{dg_n}{dx}=\frac{\lambda^2}2(n+1)(1-x)(f_{n+1}+f_n)
  -2\omega g_n
  \mp \Delta (-1)^ng_n(f_{n+1}+f_n),
\end{equation}
with the initial conditions
\begin{equation}
  f_n(0)=1,\quad g_n(0)=0.
\end{equation}
These equations are still exact. 
For large $n$ one has $g_{2n+2}\approx g_{2n}$ 
and $g_{2n+1}\approx g_{2n-1}$.
As a consequence one obtains $f_{2n+1}\approx f_{2n}$. This yields
\begin{equation}
  \omega(1-x)\frac{df_{2n}}{dx}
  \approx\omega(1-x)\frac{df_{2n+1}}{dx}
  =-g_{2n}-g_{2n+1}
\end{equation}
\begin{equation}
  2\omega(1-x)\frac{dg_{2n}}{dx}\approx2\lambda^2n(1-x)f_{2n}
  -2\omega g_{2n}
  \mp 2\Delta g_{2n}f_{2n}
\end{equation}
\begin{equation}
  2\omega(1-x)\frac{dg_{2n+1}}{dx}\approx2\lambda^2n(1-x)f_{2n}
  -2\omega g_{2n+1}
  \pm 2\Delta g_{2n+1}f_{2n}
\end{equation}
Taking the derivative of the first equation, and using the sum of the
second and the third to express the derivative of $g_{2n}+g_{2n+1}$
by $f_{2n}$ I obtain for large $n$ (i.e. $2n+1\approx 2n$)
\begin{equation}
  (1-x)\frac{d^2f_n}{dx^2}=-2\frac{\lambda^2}{\omega^2}nf_n
\end{equation}
The general solution of this differential equation can be expressed 
using Bessel functions
\begin{equation} \label{fnx}
  f_n(x)=\sqrt{1-x}\left[aJ_1\left(\frac{2\lambda}{\omega}\sqrt{n(1-x)}\right)
    +bY_1\left(\frac{2\lambda}{\omega}\sqrt{n(1-x)}\right)\right]
\end{equation}
Using the initial conditions $f_n(0)=1$ and $f'_n(0)=0$, one obtains
\begin{eqnarray}
  a&=&\pi\frac{\lambda}{\omega}\sqrt{n}
  Y_0\left(\frac{2\lambda}{\omega}\sqrt{n}\right)
  \nonumber \\
  b&=&-\pi\frac{\lambda}{\omega}\sqrt{n}
  J_0\left(\frac{2\lambda}{\omega}\sqrt{n}\right)
\end{eqnarray}
This yields the complete solution for $\epsilon_n(\ell)$. I am interested
in the limit $\ell\rightarrow\infty$, which corresponds to $x=1$. Using
the behaviour of the Bessel function for small arguments one obtains 
\begin{equation}
  f_n(1)=-J_0\left(\frac{2\lambda}{\omega}\sqrt{n}\right)
\end{equation}
This yields the final expression for the eigenvalues for large $n$
\begin{equation}\label{epsasym1}
  \epsilon_n(\ell=\infty)
  =n\omega-\frac{\lambda^2}{4\omega}
  \mp(-1)^n\frac{\Delta}2J_0\left(\frac{2\lambda}{\omega}\sqrt{n}\right)
\end{equation}
Using the asymptotic behaviour of the Bessel function, one obtains
\begin{equation} \label{epsasym2}
  \epsilon_n(\ell=\infty)
  =n\omega-\frac{\lambda^2}{4\omega}
  \mp(-1)^n\frac{\Delta}{2n^{1/4}}\sqrt{\frac{\omega}{\pi\lambda}}
  \cos\left(\frac{2\lambda}{\omega}\sqrt{n}-\frac14\pi\right)
\end{equation}
Except for a misprint (the factor $n^{-1/4}$ is missing in the last term)
this expression coincides with the 
result in \cite{GH84}. Graham {\it et al}. performed a first order
perturbational treatment in $\Delta$ and expanded the result
to obtain (\ref{epsasym2}) for large $n$. Thus, in their 
approach, the validity of (\ref{epsasym2}) is unclear.

How accurate are (\ref{epsasym1}, \ref{epsasym2})?
A first condition is obtained from 
a consistency check of the above assumption
$f_{n+1}(x)\approx f_n(x)$. This assumption must be true
for all $x$ and all $\lambda/\omega$. 
Due to $2\sqrt{n+1}\lambda/\omega\approx 2\sqrt{n}\lambda/\omega
+\lambda/(\omega\sqrt{n})$ one must have
\begin{equation} \label{condf}
  \frac{\lambda}{\omega\sqrt{n}}\ll 1
\end{equation}
But this is not the only condition one needs. 
A similar consistency check has to be done for $g_n$.
It is more complicate since it depends on $\Delta$.
Fortunately there is a more simple 
possibility to determine the range of applicability for 
(\ref{epsasym1}, \ref{epsasym2}), which is equivalent to the
consistency check proposed above. From the general
considerations in section 2 one knows that the flow equations yield
$\epsilon_n(\ell=\infty)\le\epsilon_{n+1}(\ell=\infty)$. 
This means that one must have
\begin{equation} 
  \omega>\frac{\Delta}2\abs{J_0\left(\frac{2\lambda}{\omega}\sqrt{n}\right)
    -J_0\left(\frac{2\lambda}{\omega}\sqrt{n+1}\right)}.
\end{equation}
For large $\frac{\lambda}{\omega}\sqrt{n}$ this yields the condition
\begin{equation}\label{cond}
  \frac{\Delta}{2\omega}\sqrt{\frac{\lambda}{\pi\omega}}\frac1{n^{3/4}}<1
\end{equation}
(\ref{condf}) and (\ref{cond}) show that within a wide range of parameters 
(\ref{epsasym1}, \ref{epsasym2}) are applicable. But one should
be careful using (\ref{epsasym2}) for small $\lambda$. (\ref{epsasym2})
has a relative error $O(\frac{\omega}{2\lambda\sqrt{n}})$ compared
to (\ref{epsasym1}) so that for small $\lambda$ (\ref{epsasym1})
yields better results. For $\lambda=0$, (\ref{epsasym1})
yields the exact solution (which is of course trivial),
whereas (\ref{epsasym2}) is not defined.
Only in the limit $\lambda\rightarrow 0$
and $\Delta\lambda/\omega^2={\rm const}$ it is not possible
to apply (\ref{epsasym1}). But this regime can be treated with
perturbation theory for small $\lambda$.

The accuracy of (\ref{epsasym1}) is very high, even for small $n$ and
large $\Delta$. It can be tested numerically if one compares (\ref{epsasym1}) 
with the exact numerical solution of the flow equations. This is done
in figure 1. 
\begin{figure}[ht]
\leavevmode
\centering
\epsfig{file=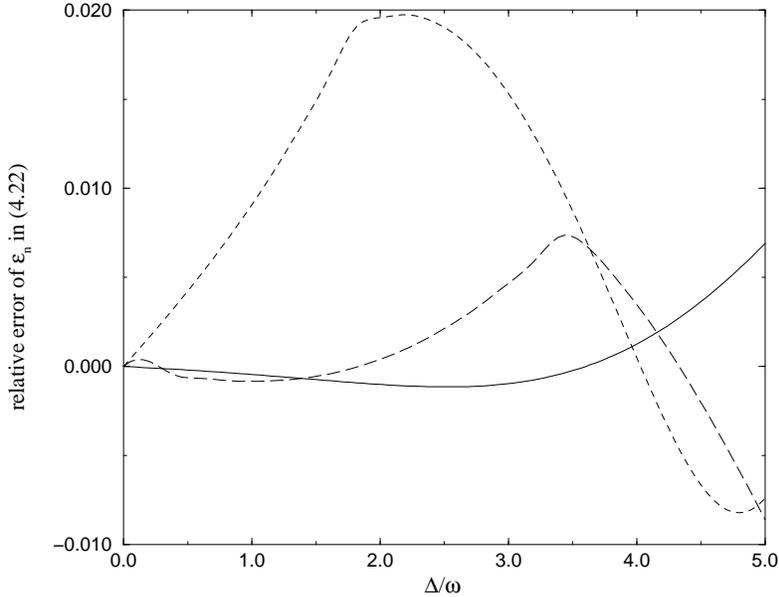,scale=.5,angle=270} 
\caption{The relative error of the asymptotic expression for the
eigenvalue compared to the exact result as a function of
$\Delta/\omega$ for $\lambda/\omega=4.0$. The short dashed line
is $n=10$, the long dashed line $n=15$ and the solid line $n=20$.} 
\end{figure}
We show the relative error of the asymptotic formula 
(\ref{epsasym1}) plotted as a function of $\Delta/\omega$. The coupling
is $\lambda/\omega=4.0$. For small $n$ the error is relatively large
(up to 2\% for $n=10$), whereas it is smaller for larger $n$. The parameters
are chosen so that for $n=10$ the left hand side of (\ref{condf}) equals
1.3 and is therefore too large. 
The left hand side of (\ref{cond}) equals 0.5 
for the worst case shown, i.e. 
$n=10$ and $\Delta=5\omega$. A relative error less than 0.1\% 
is obtained if the left hand side of (\ref{cond}) is less
than 0.1 and for $n\gtrapprox \lambda^2/\omega^2$.

The main result of this section is clearly the expression 
for $\epsilon_n(\ell=\infty)$ given above. But one does not
only know the eigenvalues of the Hamiltonian, the
complete flow (\ref{fnx}) is known as well. This allows
to reconstruct the continuous unitary transformation that
diagonalizes the Hamiltonian. Therefore it is even possible
to calculate other observables in the transformed basis.
In principle it is possible to combine this result with the
results of Richter \cite{Richter97}, which are very accurate for
small $n$. Then one should be able to obtain accurate 
values for dynamical correlation functions of the model.
But the goal of the present calculations was to show that the new flow
equations proposed in this paper are indeed useful if one
wants to deal with band-diagonal matrices or Hamiltonians. 
Therefore I do not follow this line.

\Section{Conclusions}

The flow equations proposed in this paper seem to be useful
if one wants to solve a given model. The two examples showed that
solutions can be obtained at least approximatively or in some limits.
Some of the advantages of the new flow equations have already been
pointed out: 1. The matrix is finally in a diagonal form, even if
degeneracies occur. 2. If the Hamiltonian has initially a banded
structure, this form is preserved. 3. The asymptotic behaviour of
the equations is known. Another advantage is that with the new
flow equations one can obtain accurate results for higher eigenvalues
as well. But there are also some disadvantages, which may be
important in other systems:
\begin{enumerate}
\item Although asymptotically off--diagonal matrix elements
decay faster if the difference of the corresponding diagonal
matrix elements is larger, the new generator does not separate
different energy scales automatically. For small $\ell$ it is
not guaranteed that off--diagonal matrix elements corresponding
to large energy differences decay fast. Therefore the new
flow equations do not provide a renormalization scheme.
\item An important property of a Hamiltonian one wants to
solve using the new flow equations is that it must have a pure
point spectrum. Hamiltonians with continuous spectra cannot
be treated that way. It is for instance not possible to
apply the new generator to a dissipative quantum system.
\item If the diagonal matrix elements of the initial Hamiltonian
are not in the correct order (i.e. $h_{nn}\le h_{mm}$ if $n<m$)
the flow equations will reorder the diagonal matrix elements.
This may cause a problem, because in such a case the analytical
treatment of the flow equations becomes more difficult. 
\end{enumerate}
The first two points mentioned above show that the new generator
cannot be applied successfully to the kind of problems that
have been treated so far using flow equations. In all these
problems one has several different energy scales and continuous
spectra, and one needs a renormalization scheme to obtain
useful results. In this sense the new generator provides a complementary
set of flow equations. It can be applied to problems that cannot
be treated with the original flow equations, but problems that
can be treated with the original flow equations are not within
the range of possible applications of the new scheme.

What kind of problems can be treated with the new flow equations?
It became already clear that Hamiltonians with a banded structure
are good candidates. But even for Hamiltonians without such a
structure the present approach may be useful. If one has for
instance a problem for which all diagonal matrix elements are of the same
order of magnitude, the usual flow equations are difficult to apply.
The reason is that differences of diagonal matrix elements are small
and that therefore the flow is very slow. This does not happen with
the new flow equations, since the flow of the diagonal matrix elements
is only determined by the magnitude of the off-diagonal matrix elements.
A class of possible candidates are therefore disordered systems.

In general one can say that the new flow equations can be applied
to single or few particle systems. It may therefore be useful
in nuclear or atomic physics.

\vspace{1cm}
\subsection*{Acknowledgement}
I wish to thank F. Wegner, H.J. Pirner, J. Richter, H. Kunz, and M.B. Cibils
for useful discussions.

\clearpage
%%%%%%%%%%%%%%%%%%%%%%%%%%%%%%%%

%%%%%%%%%%%%%%%%%%%%%%%%%%%%%%%%%%%%%%%%

\begin{thebibliography}{99}
%%%%%%%%%%%%%%%%%%%%%%%%%%%%%%%%
\bibitem{Wegner94} F. Wegner: Ann. Phys. (Leipzig) {\bf 3}, 77 (1994).

\bibitem{SK97a} S. K. Kehrein, and A. Mielke:
  Ann. Phys. (Leipzig) {\bf 6}, 90 (1997).

\bibitem{SK97b} S. K. Kehrein, and A. Mielke:
  to appear in J. Stat. Phys. (1998).

\bibitem{LW96} P. Lenz, and F. Wegner:
  Nucl. Phys. B {\bf 482} [FS], 693 (1996).

\bibitem{Mielke97a} A. Mielke:
  Ann. Phys.  (Leipzig) {\bf 6}, 215 (1997).

\bibitem{Mielke97b} A. Mielke:
  Europhys. Lett. {\bf 40}, 195 (1997).

\bibitem{GW94} S. D. G{\l}azek, and K. G. Wilson: 
  Phys. Rev. D {\bf 49}, 4214 (1994).

\bibitem{Wilson94} K.G. Wilson,  T.S. Walhout, A. Harindranath,
  W.M. Zhang, R.J. Perry, and  S.D. G{\l}azek: 
  Phys. Rev. D {\bf 49}, 6720 (1994).

\bibitem{Brisudova97} M.M. Brisudov\'a M.M., R.J. Perry R.J., and  K.G. Wilson:
  Phys. Rev. Lett. {\bf 78}, 1227 (1997).

\bibitem{Richter97} J. Richter: Diploma thesis, Heidelberg 1997.
J. Richter, and A. Mielke, in preparation.

\bibitem{Lipkin65} H.J. Lipkin, N. Meshkov, and A.J. Glich: 
  Nucl. Phys. {\bf A62}, 188, 199, 211 (1965).

\bibitem{PF98} H.J.Pirner, and B. Friman: 
  'Hamiltonian Flow Equations for the Lipkin Model.' preprint 1998.

\bibitem{DS90} J. Dukelsky, and P. Schuck: 
  Nucl. Phys. {\bf A512}, 466 (1990).

\bibitem{GH84} R. Graham, and M. Höhnerbach: 
  Z. Phys. B {\bf 57}, 233 (1984).

\bibitem{Cibils91} M.B. Cibils, Y. Cuche, V. Marvulle, W.F. Wreszinski, 
  J.-P. Amiet, and H. Beck: 
  J. Phys. A {\bf 24}, 1661 (1991).

\bibitem{Cibils92} M.B. Cibils, Y. Cuche, P.Leboeuf,  and W.F. Wreszinski:
  Phys. Rev. A{\bf 46}, 4560 (1992).






\end{thebibliography}
\end{document}